\documentclass[journal=jacsat,manuscript=article]{achemso}

\usepackage{chemformula} % Formula subscripts using \ch{}
\usepackage[T1]{fontenc} % Use modern font encodings

\usepackage{times}
\usepackage[english]{babel}
\usepackage{bm}
\usepackage{amsmath}
\usepackage{amssymb}
\usepackage{lipsum}
\usepackage{graphicx}
\usepackage{diagbox}
\usepackage{multirow}
\usepackage{threeparttable}
\usepackage{array}
\author{Aitor Garcia-Ruiz}
\affiliation{School of Physics and Astronomy, University of Manchester, Oxford Road, Manchester, M13 9PL, UK}
\alsoaffiliation{National Graphene Institute, University of Manchester, Oxford Road, Manchester, M13 9PL, UK}
\email{altor.garcia-ruiz@manchester.ac.uk}
\author{Vladimir Enaldiev}
\affiliation{School of Physics and Astronomy, University of Manchester, Oxford Road, Manchester, M13 9PL, UK}
\alsoaffiliation{National Graphene Institute, University of Manchester, Oxford Road, Manchester, M13 9PL, UK}
\author{Andrew McEllistrim}
\affiliation{School of Physics and Astronomy, University of Manchester, Oxford Road, Manchester, M13 9PL, UK}
\alsoaffiliation{National Graphene Institute, University of Manchester, Oxford Road, Manchester, M13 9PL, UK}
\author{Vladimir I. Fal'ko}
\affiliation{School of Physics and Astronomy, University of Manchester, Oxford Road, Manchester, M13 9PL, UK}
\alsoaffiliation
{National Graphene Institute, University of Manchester, Oxford Road, Manchester, M13 9PL, UK}
\alsoaffiliation{Henry Royce Institute for Advanced Materials, University of Manchester, Oxford Road, Manchester, M13 9PL, UK}

\title{Mixed-stacking few-layer graphene as an elemental weak ferroelectric material}

\begin{document}

%\begin{tocentry}
%\end{tocentry}

\begin{abstract}
Ferroelectricity \cite{Valasek_Piezo-electric_1921} - a spontaneous formation of electric polarisation - is a solid state phenomenon, usually, associated with ionic compounds or complex materials. Here we show that, atypically for elemental solids, few-layer graphenes can host an equilibrium out-of-plane electric polarisation, switchable by sliding the constituent graphene sheets. The systems hosting such effect include mixed-stacking tetralayers and thicker (5-9 layers) rhombohedral graphitic films with a twin boundary in the middle of a flake. The predicted electric polarisation would also appear in marginally (small-angle) twisted few-layer flakes, where lattice reconstruction would give rise to networks of mesoscale domains with alternating value and sign of out-of-plane polarisation.
\end{abstract}
$\mathbf{Keywords}$: Ferroelectricity; graphene; rhombohedral graphite; twin boundary; twistronics; screening


Ferroelectricity, a spontaneous electric polarisation in absence of an external electric field, is a phenonmenon observed and thoroughly investigated in a broad range of solids \cite{LinesGlass1977,Uchido_Ferroelectric_2018,BainChand_Ferroelectric_2010}. Caused by charge transfer between constituent atoms in a unit cell, up to now, ferroelectric polarisation was observed only in chemically complex compounds, such as Rochelle salts \cite{Valasek_Piezo-electric_1921}, $\mathrm{Pb[Zr_xTi_{1-x}]O_3}$ \cite{Jaffe_Piezoelectric_1954}, $\mathrm{BaTiO_3}$ \cite{Megaw_Origin_1952}, etc... \cite{Chang_Discovery_2017,De_la_Barrera2021,Sharma2019,Fei_2018,Weston2022}. Here, we show that there is one exception from this common rule, namely, several structural allotropes of multilayer graphene.

The above statement is based on a theoretical study of various multilayer graphene structures with different stacking orders which, as an elemental material, is non-polar: its intralayer bonding is dominantly covalent, whereas the interlayer adhesion has a van der Waals nature. There are two commonly studied multilayer graphene systems: thin films of Bernal graphite \cite{Bernal_structure_1924,freise_structure_1962} and rhombohedral (ABC) graphite \cite{latychevskaia_stacking_2018}. Those two possess inversion symmetry, or $z\to-z$, respectively, which prohibit spontaneous ferroelectric polarisation. However, graphitic films with mixed Bernal and rhombohedral stackings lack such symmetries, removing constraints on the formation of out-of plane electric dipole. One example of such a structure is a thin film of rhombohedral graphite with a twin boundary inside it \cite{Aitor_electronic_2022}, which consists of two ABC parts with $n$ and $m$ layers ($n\neq m$), rotated by $180^\circ$ with respect to each other, held together by an ABA-trilayer (see Fig. 1).

\textcolor{black}{In the following section, we study $n$ABA$m$, with $n\neq m$, and demonstrate that they exhibit a weak spontaneous out-of-plane electric polarisation, $P_z$, at zero doping. We determine the roles of parameters in the full multi-parameter Slonczewski-Weiss-McClure (SWMcC) Hamiltonian that are related to the asymmetries that permit a finite $P_z$. Importantly, we find that self-consistently implemented screening of internal electric field not only strongly reduces $P_z$ magnitude, but also changes its sign for some structures, which is a result of different inter-layer charge redistributions, caused individually by each symmetry-breaking term in the bare SWMcC Hamiltonian. To mention, for tetralayers, the known values and signs of SWMcC parameters\cite{yin_dimensional_2019} are such that the largest contributions they produced individually mostly cancel each other, making the result sensitive to the precise choice of parameters. At the same time, for marginally twisted tetralayers, where lattice relaxation leads to the formation of equilibrium stacking domains (where polarisation is small), the domain walls host seeds of different stacking arrangements for which such cancellation does not occur leading to an order of magnitude larger local $P_z$ values.}
\begin{figure}
\begin{center}
\includegraphics[width=0.9\columnwidth]{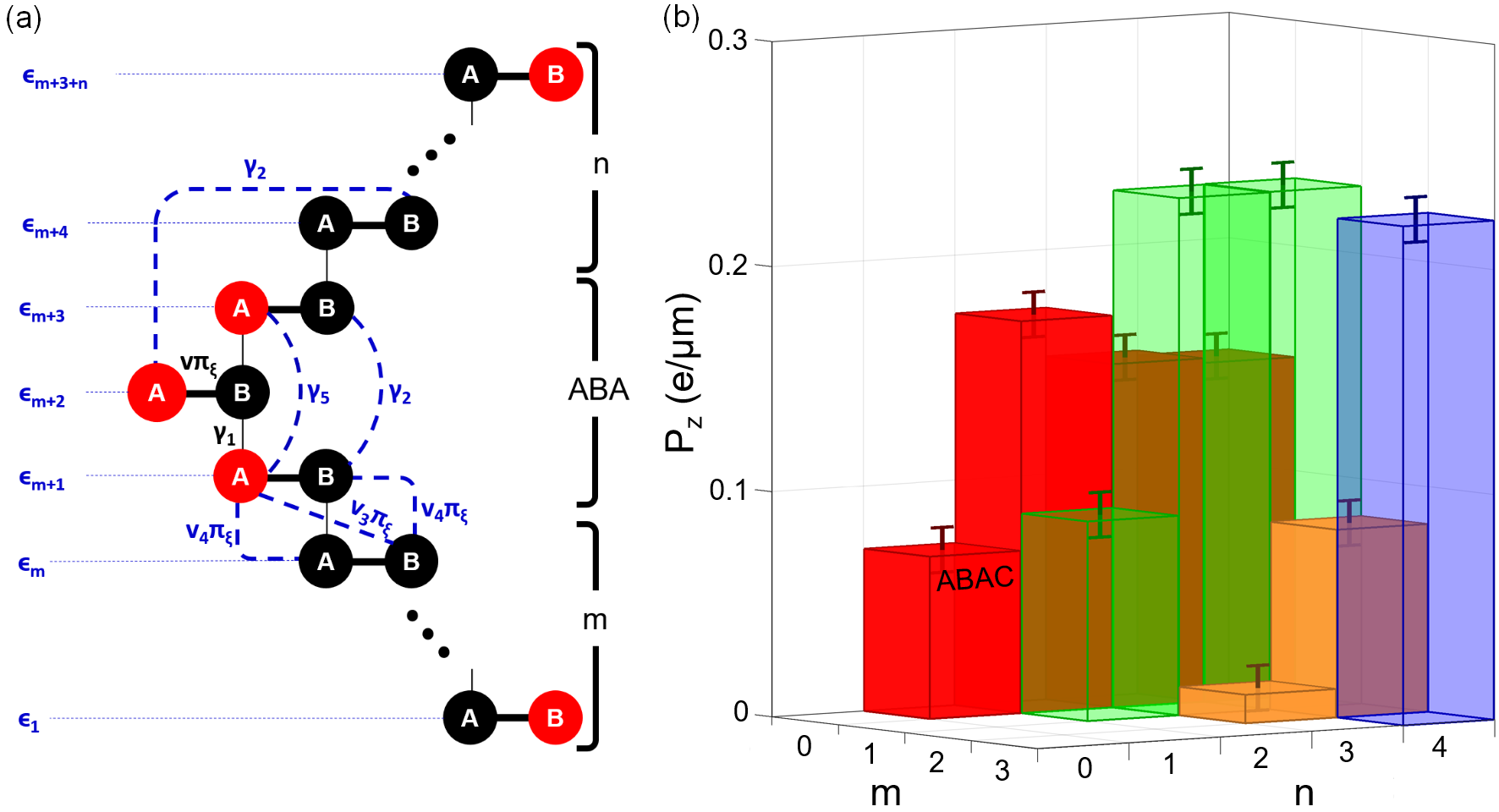}
\caption{ Sketch of a twinned rhombohedral graphitic film, {highlighting in dashed lines the SWMcC couplings, with a radius proportional to the number of dimmer bonds and showing in red} the low-energy orbitals (three non-dimer bonds and the anti-symmetric combination of A-sublattice orbitals of the layers adjacent to the twin boundary). The right panel show the dependence of the ferroelectric polarisation density as a function of thicknesses $n$ and $m$ of twinned layers. 
\label{Fig1}}
\end{center}
\end{figure}

\textcolor{black}{Below,} we use a full Slonczewski-Weiss-McClure (SWMcC) model of graphite films discussed in Ref. \cite{Aitor_electronic_2022}. This model accounts for couplings sketched in Fig. 1, which include both the closest and next-neighbour hoppings, and it is implemented in the framework of a hybrid $\bm{k}\cdot\bm{p}$-tight-binding model \cite{Garcia-Ruiz2021}, using a Hamiltonian, $\mathcal{H}$, specified in the Supplementary Materials. \textcolor{black}{The diagonalisation of $\mathcal{H}$ gives the dispersions, $\varepsilon_{\beta}(\boldsymbol{p})$, and wavefunctions, $\psi_{\beta,\boldsymbol{p}}^{\alpha_L}$, of bands $\beta$ in the multilayer, which we use to compute both on-layer electron densities, $n_L$, and electric polarisation:
\begin{align}\label{Eq:Pz_densities}
P_z=ed
\sum_{L=1}^{n+m+3}Ln_L;\quad
n_{L}=4
\sum_{\beta,\alpha_L,\boldsymbol{p}}
\left[
\left|
\psi_{\beta,\boldsymbol{p}}^{\alpha_L}
\right|^2
\Theta[E_\mathrm{F}-\varepsilon_{\beta}(\boldsymbol{p})]-\frac{1}{4}
\right].
\end{align}}
\textcolor{black}{Here $e$ is the electron charge, $d\approx3.35$~\AA ~the interlayer distance, and $``L"$ the layer index, and $\alpha_L=A_L,B_L$ sublattice indices. When computing the electron densities, $n_L$, we take account for the screening the potential they generate on each individual layer, as
\begin{align}\label{Eq:Ediffs}
\epsilon_i-\epsilon_{i-1}=
\frac{e^2d}{2\varepsilon_0}
\left[
(n_i-n_{i-1})
\frac{1+\varepsilon_z}{2\varepsilon_z}+
\sum_{j>i}
\frac{n_j}{\varepsilon_z}-
\sum_{j'<i-1}
\frac{n_{j'}}{\varepsilon_z}
\right],
\end{align}
where $\varepsilon_z\approx2.6$ is an effective out-of-plane dielectric permittivity of graphene stacks determined by the polarizability of carbon atoms \cite{Slizovskiy_dielectric_2021}.}

Among all SWMcC couplings, next-neighbour hoppings $\gamma_2$, $\gamma_5$ and $v_4$, together with the energy difference between dimer and non-dimer sites, $\Delta'$, are most important for a non-zero value of the ferroelectric polarisation, $P_z$, to form across the structure. This is because such couplings break a hidden electron-hole symmetry\footnote{Applying  $\mathcal{U}$ to Hamiltonian,  $\mathcal{U}\mathcal{H}\mathcal{U}^\dagger$, we change signs all of its matrix elements except those on the diagonal of matrices $H_{\rm g}^{s,b}$, $V$, $W$, and $\tilde{W}$ being proportional to either of the parameters $v_4, \Delta', \gamma_2, \gamma_5$. Therefore, for $v_4=\Delta'=\gamma_2=\gamma_5=0$ Hamiltonian possesses the hidden electron-hole symmetry $\mathcal{U}\mathcal{H}\mathcal{U}^\dagger=-\mathcal{H}$, which results in $P_z=0$. This is because at zero doping polarisations given by electron and hole parts of spectrum should be opposite, whereas the electron-hole symmetry equalizes them requiring zero value for both.}, $\mathcal{U}\mathcal{H}\mathcal{U}^\dagger=-\mathcal{H}$ ($\mathcal{U}={\hat{1}}_{n+3+m}\otimes\sigma_z$ is a unitary matrix equal to direct product of \mbox{$n+3+m$}-rank identity and third Pauli matrices, respectively), characteristic of the 'reduced' models of graphenes limited to the closest neighbour couplings only. \textcolor{black}{To mention, the on-layer potentials in Eq. (\ref{Eq:Ediffs}) induced by the screening can also break all symmetries, which will be important for understanding the result of self-consistent analysis.}

\textcolor{black}{For a detailed quantitative self-consistent analysis of spontaneous electric polarisation $P_z$, we implement the following steps. First, we compute the electric polarisation induced by one of the four symmetry-breaking terms in SWMcC Hamiltonian for each of the considered structures, both with and without self-consistent implementation of screening. Then, we check that the cumulative effect of all the terms in the full SWMcC model can be approximation as a sum of the individual symmetry-breaking contributions as}
\begin{align} \label{Eq:Pz}
P_z=
\left(
\mathcal{X}_{4}\frac{v_4}{v}\gamma_1+
\mathcal{X}_{D}\Delta'+
\mathcal{X}_{2}\gamma_2+
\mathcal{X}_{5}\gamma_5
\right)
\frac{|\gamma_1|}{\hbar^2v^2}
\cdot
ed.
\end{align} 
\textcolor{black}{Here, $\mathcal{X}_{4,D,2,5}$ are dimensionless factors which values for the ABCB tetralayer are listed in Table \ref{Tab:X}.}  

\begin{table}[t]
\caption{\textcolor{black}{Numerical values for the parameters $\mathcal{X}$ in Eq. (\ref{Eq:Pz}), and their contribution towards the polarisation, with and without including the screening effects of on-layer charge redistribution. In the last four rows we also include the on-layer potentials. To evaluate the dependence with respect to each parameter, we use a Hamiltonian model where all other SWMcC parameters were set to zero except $v=1.02\cdot10^6$ m/s and $\gamma_1=390$ meV.\label{Tab:X}}}
		\begin{tabular}{|c||c||ccc||c|c|c|c|}
        \hline
			\hline
   
		  & $\frac{P_z^u}{\mathrm{e/\mu m}}$
            & 
            & Coeffs.
            &  
            &  $\frac{P_z}{\mathrm{e/\mu m}}$ 
            & $\frac{\epsilon_2-\epsilon_1}{\mathrm{meV}}$
            & $\frac{\epsilon_3-\epsilon_2}{\mathrm{meV}}$
            & $\frac{\epsilon_4-\epsilon_3}{\mathrm{meV}}$
            \\
			\hline 
		Total
            &0.50
            &
            &
            &
            &-0.07
            &
            &
            &\\
			\hline 
			\hline 
		  $v_4=0.022v$
            &0.09  
            &0.036
            &$\mathcal{X}_4$
            &0.009
            &0.02
            &-1.332
            &1.323
            &-0.418
            \\
			\hline 
		  $\Delta'=25$ meV
            &-0.33  
            &-0.045
            &$\mathcal{X}_D$
            &-0.032
            &-0.23
            &-5.607
            & 5.631
            &0.716
            \\
			\hline 
		  $\gamma_2=-17$ meV
            &0.28  
            &-0.058
            &$\mathcal{X}_2$
            &0.016
            &-0.08
            &1.403
            &-1.959
            &2.158\\
			\hline 
		  $\gamma_5=38$ meV
            &0.45 
            &0.041
            &$\mathcal{X}_5$
            &0.019
            &0.21
            &1.498
            &-1.567
            &-1.404\\
        \hline
        &
        \multicolumn{1}{c}{unscreened}
        &&&
        \multicolumn{4}{c}{self-consistently screened}
        &\\
			\hline 
			\hline
		\end{tabular}
\end{table}

\textcolor{black}{Accounting for the charge redistribution self-consistently is an important part of the presented calculations. Taking into account redistribution of charges produced by screening of electric fields attributed to polarisation not only reduces the value of $P_z$ by one order of magnitude, but also can change the sign of some of the individual symmetry-breaking contributions, when compared with the unscreened case. We investigate numerically this effect by comparing the values for $\mathcal{X}$ in ABCB graphenes, with and without the self-consistent implementation of screening in Table \ref{Tab:X}. In absence of screening ($\epsilon_i=0$), we obtain a value of $P_z\approx0.5~\mathrm{e/\mu m}$, whereas implementing screening changes this values to $P_z=-0.07~\mathrm{e/\mu m}$. This is because the two largest contributions to the ferroelectric polarisation, coming from $\mathcal{X}_D$ and $\mathcal{X}_D$, have opposite signs and nearly cancel each other for the choice of SWMcC based on Ref. \cite{yin_dimensional_2019}, leaving $\gamma_2$ to define the value of $P_z$, so that the overall result changes sign after the implementation of screening. We suggest that such a sensitivity of the computed values of $P_z$ to the input SWMcC parameters may be used to further narrow down their choices by comparing the computed $P_z$ with the experimentally measured polarisation of tetralayers. We find that screening is equally important for the analysis of polarisation in thicker films, though the above-mentioned cancellation does not occur for all thicknesses. In Fig. \ref{Fig1}, we show the computed values of $P_z$ in various nABAm structures with $m>n$ (the mirror-symmetric mABAn configurations have the same magnitude of $P_z$ but opposite sign). }

% Thickness-dependence of the ferroelectric polarisation, shown in Fig. 1, is determined by the interplay of two contributions. The first one, produced by low-energy states localised at the twin boundary and non-dimer sites at the surfaces (highlighted in red in Fig. 1), is a major contribution for structures with large twin plane-surface separation (i.e. large $n$, $m$, $n\neq m$) that suppresses the state hybridization and leads to flat bands around Fermi-level. This is confirmed by saturation of momentum integral in Eq. \eqref{Eq:Pz_densities} for momentum cut-offs exceeding range of flat dispersion (see Supplementary Materials). In contrast, for structures with $n=0$ and $m\geq1$ (or $m=0$ and $n\geq 1$) strong hybridization between twin boundary and surface states makes the low-energy state dispersive, increasing saturation momentum cut-offs and reducing their contribution to polarisation \eqref{Eq:Pz_densities}. Moreover, for these structures there is another competing contribution resulted from all the other filled states under Fermi-level having the opposite-sign with respect to the first one. Therefore, sum of the two contributions reduces the magnitude of total polarisation for $m$ABA (ABA$n$) structures in comparison with $m$ABA$n$.    

While the above-described twinned rhombohedral graphitic films may appear naturally among {flakes} exfoliated from bulk material, one can try to build weakly ferroelectric multilayers by a twistronic assembly of thinner flakes \cite{Kim_vdW_2016,caldwell_technique_2010,Kim_tunable_2017,xu_tunable_2021,Tomic_Scattering_2022}. A small misalignement of crystalline axes of the assembled graphene flakes, unavoidable in the mechanical transfer, leads to a long period variation of stacking, known as moir\'{e} structure, which upon the lattice relaxation (promoting energetically preferred local stacking order) results in a network of domains with AB and BA interlayer preferences at the twisted interface. 

\begin{figure}
\begin{center}
\includegraphics[width=0.6\columnwidth]{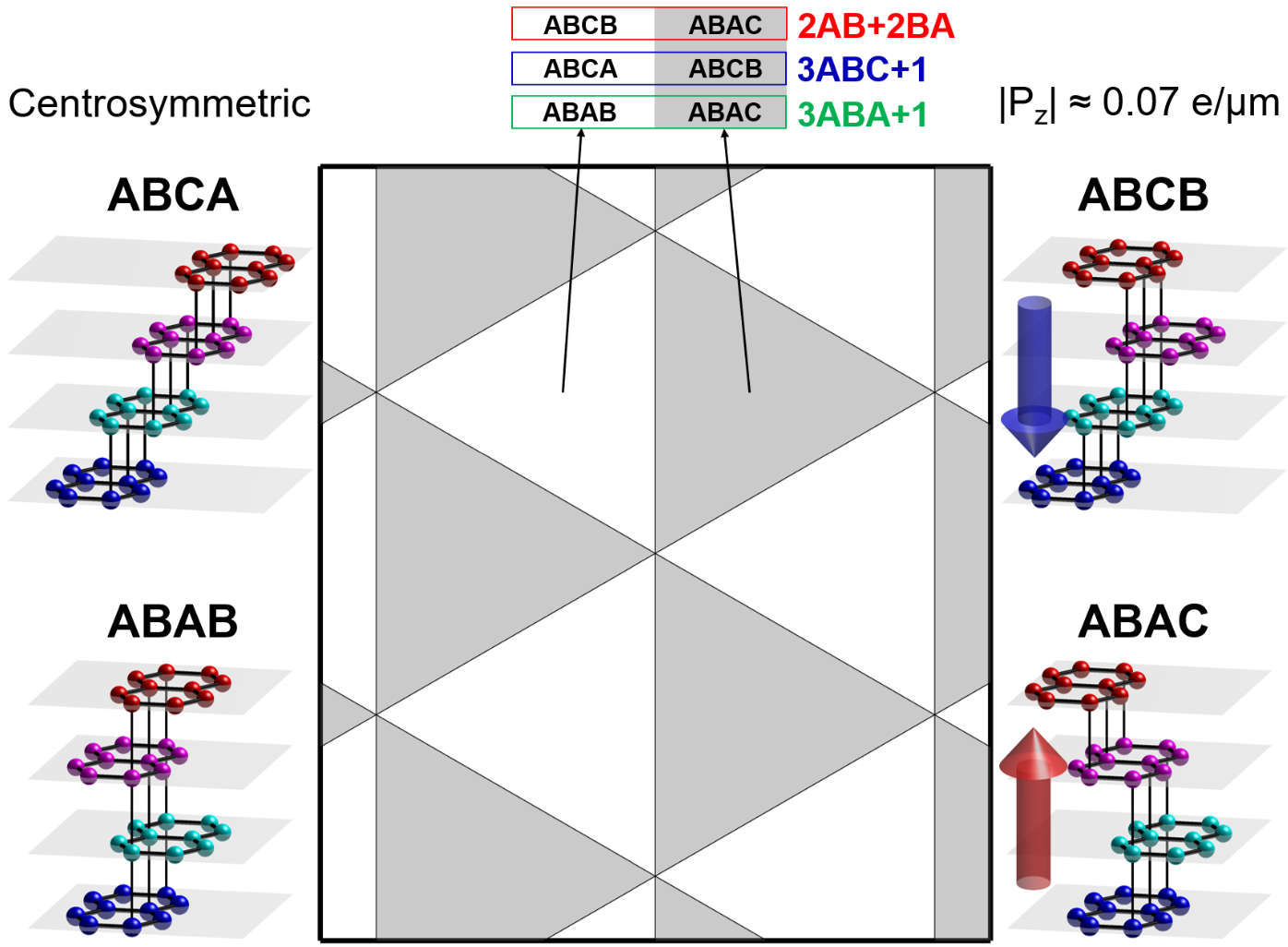}
\caption{ An array of triangular domains in a marginally twisted tetralayer graphene (2AB+2BA, 3ABA+1 and 3ABC+1), which we expect to form by lattice reconstruction of a long-period moir\'e superlattice. Domain stackings with and without ferroelectric charge transfer for various twisted tetralayer structures are shown on the right and left hand side sketches, respectively.
\label{Fig2}}
\end{center}
\end{figure}

Here, we consider structures with graphene monolayer small-angle-twisted with respect to Bernal and rhombohedral trilayer graphene (3ABA+1 and 3ABC+1, respectively) and a twisted double-bilayer (2AB+2BA). At the twisted interface, moir\'e superlattice undergoes lattice reconstruction promoting formation of triangular domain arrays with local AB and BA stackings, shown in Fig. 2, which are separated by network of domain walls (DWs). We model lattice reconstruction in the structures following an approach of Refs. \cite{Nguyen_2017,EnaldievPRL2020}, find local deformations of lattices in top/bottom flakes, $\bm{u}_{t/b}$, that minimise (via domain formation) a sum of elastic and interlayer adhesion energies (see Supplementary Materials). 

Then, we find ferroelectric polarisation of small-angle twisted structures using local stacking approximation. That is, we calculate the polarisation based on a Hamiltonian shown in Methods, describing aligned tetralayer films with {an} in-plane offset $\bm{r}_0=\theta\hat{z}\times\boldsymbol{r}+\boldsymbol{u}_t-\boldsymbol{u}_b$ between constituent parts. As a reference, we set $\bm{r}_0=0$ for ABBA-stacking in 2AB+2BA, ABAA-stacking in 3ABA+1, and ABCC-stacking in 3ABC+1 areas. Having calculated $P_z$ with a self-consistent analysis of screening and on a sufficiently dense grid of interlayer offsets $\bm{r}_0$, we interpolate it in the form of a harmonic series,
\begin{align}
\label{Eq:Pz_Fourier}
&P_z(\bm{r}_0)=
\sum_{{\Lambda}=0}^{+\infty}
\sum_{j=0}^2
\sum_{s=\pm}\left[
A_\Lambda
\sin\left(
\bm{G}^{j,s}_{ {\Lambda}}\bm{r}_0\right)+
B_\Lambda
\cos\left(s
\bm{G}^{j,s}_{ {\Lambda}}\bm{r}_0\right)\right],\nonumber\\
&\bm{G}^{j,s}_{{\Lambda}}=
\frac{4\pi}{a\sqrt{3}}
\begin{pmatrix}
\lambda_1\sin\left(
\frac{2\pi}{3}(j+\frac{s}{2})
\right) - 
\lambda_2\sin\left(\frac{2\pi}{3}j\right) \\
-\lambda_1\cos\left(
\frac{2\pi}{3}(j+\frac{s}{2})
\right) + \lambda_2\cos\left(\frac{2\pi}{3}j
\right)
\end{pmatrix} {,}
\end{align}
{where $\Lambda$ labels reciprocal lattice vectors of graphene in the ascending order of the lengths (see Table \ref{Tab:Fourier})}. Note that mirror reflection of AB and BA bilayers in 2AB+2BA films translates to symmetry properties of $P_z(\bm{r}_0)$. In particular, for a reference frame with $x$-axis along zigzag and $y$-axis along armchair, $P_z(x_0,y_0)=P_z(-x_0,y_0)$, as zigzag axis is perpendicular to vertical mirror symmetry plane conserving $P_z$, and, $P_z(x_0,y_0)=-P_z(x_0,-y_0)$, as mirror reflection in armchair axis is equivalent to exchange of AB $\leftrightarrow$ BA stackings in each bilayer and consequently leads to inversion of $P_z$. This requires $B_\Lambda=0$ in \eqref{Eq:Pz_Fourier} for 2AB+2BA films, whereas the other Fourier coefficients are listed in Table \ref{Tab:Fourier}.

In Fig. 3, we illustrate the resulting real space distributions of the ferroelectric polarisation in a representative region of moir\'e superlattice for each of the considered tetralayer films. The mirror reflection in armchair axes prescribe zero polarisation for DWs in 2AB+2BA films, shown in Fig. 2, oriented along armchair crystallographic axes and separating domains with opposite polarisation \textcolor{black}{$P_z=\pm0.07e\mathrm{/\mu m}$}. This would correspond to the transfer of $\sim 10^{11}\,\mathrm{cm}^{-2}$ electron density between the tetralayer film surfaces. Note that close to domain corners polarisation \textcolor{black}{reverses sign and its magnitude} become four times higher than those in the middle of domains.

\begin{figure*}
\begin{center}
\includegraphics[width=1\columnwidth]{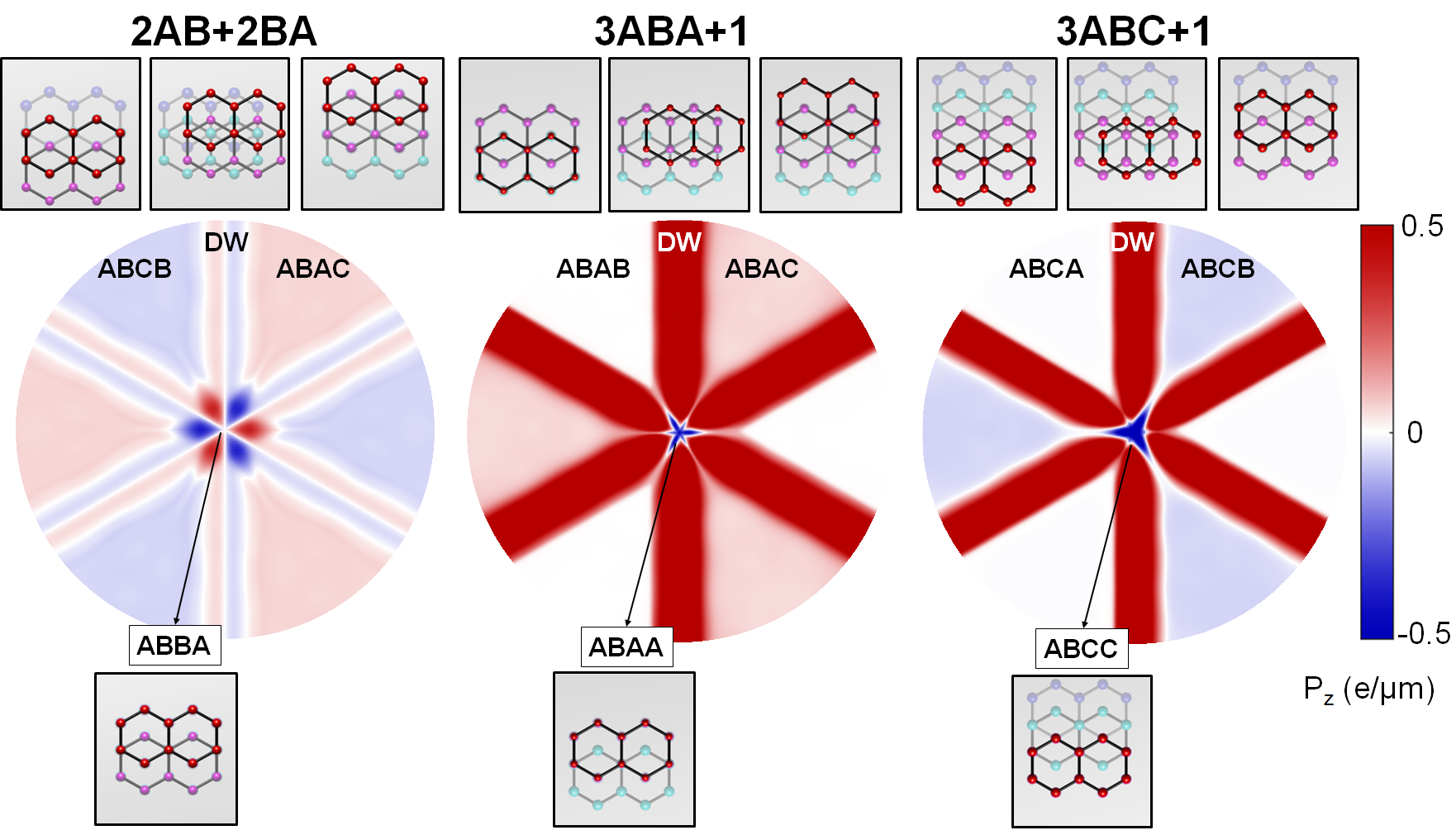}
\caption{ Ferroelectric polarisation around a circular area of 1 $\mathrm{\mu m}$ of diameter, centred at the intercross of triangular networks of domain walls, shown in Fig. \ref{Fig2}. The inset represent the top view of the local stacking configuration at the domains, the domain walls and the centre of the intercross of triangular network of domain walls. 
\label{Fig3}}
\end{center}
\end{figure*}
DW intercrossings possessing opposite polarisation compared to the inner part of DW
In contrast, supercells of 3ABA+1 (3ABC+1) twisted tetralayers  contain one inversion symmetric domain, ABAB (ABCA) (see Fig. 2), where ferroelectric polarisation is forbidden by symmetry and another, ABAC (ABCB) domain (as in 2AB+2BA films), which carries an out-of-plane polarisation. We note that in 3ABA+1 (3ABC+1) tetralayers DWs and their intercrossings host even an order of magnitude stronger local vertical charge transfer than in the inner areas of the domains.  
\begin{table}[t]
\caption{Coefficients of the Fourier expansion in Eq. (\ref{Eq:Pz_Fourier}) for the three stacking configurations analysed.\label{Tab:Fourier}}
		\begin{tabular}{l|c|c|c|c|c|c}
        \hline
        \multicolumn{2}{c|}{ }&
        AB+BA&
        \multicolumn{2}{c}{ABA+1}
        &
        \multicolumn{2}{|c}{ABA+1}\\
			\hline
			$\Lambda$ &  $\{\lambda_1,\lambda_2\}$  & $A_\Lambda$ ($e/\mathrm{\mu m}$) & $A_\Lambda$ ($e/\mathrm{\mu m}$) & $B_\Lambda$ ($e/\mathrm{\mu m}$)& $A_\Lambda$ ($e/\mathrm{\mu m}$) & $B_\Lambda$ ($e/\mathrm{\mu m}$)\\
			\hline 
			0  & $\{0,0\}$ &-       &-       & 0.0384 &-       & 0.0243  \\
			1  & $\{1,0\}$ & 0.0626 &-0.0293 &-0.1231 & 0.0488 &-0.1599  \\
			2  & $\{2,1\}$ & 0      & 0      &-0.0917 & 0      &-0.1015  \\
			3  & $\{2,0\}$ & 0.0847 &-0.0320 & 0.0927 & 0.0621 & 0.0946  \\
			4  & $\{3,1\}$ & 0.0147 &-0.0045 &-0.0319 & 0.0096 &-0.0369  \\
			5  & $\{3,0\}$ & 0      &-0.0067 & 0.0307 & 0.0015 & 0.0140  \\
			6  & $\{4,2\}$ & 0      & 0      &-0.0335 & 0      &-0.0208  \\
			7  & $\{4,1\}$ & 0.0118 &-0.0039 & 0.0151 & 0.0052 & 0.0083  \\
			8  & $\{4,0\}$ & 0.0076 &-0.0057 & 0.0024 & 0.0102 &-0.0036  \\
			9  & $\{5,2\}$ & 0.0015 & 0.0013 &-0.0130 &-0.0021 &-0.0118  \\
		    10 & $\{5,1\}$ &-0.0002 &-0.0017 & 0.0189 & 0.0001 & 0.0013  \\
	        11 & $\{5,0\}$ & 0.0062 &-0.0025 & 0.0158 & 0.0032 & 0.0009  \\
		    12 & $\{6,3\}$ & 0      & 0      &-0.0151 & 0      &-0.0060  \\
 	        13 & $\{6,2\}$ & 0.0013 & 0.0004 & 0.0033 &-0.0005 &-0.0002  \\
		    14 & $\{6,1\}$ & 0.0035 &-0.0025 & 0.0011 & 0.0041 &-0.0014  \\
		    15 & $\{6,0\}$ &-0.0004 &-0.0001 & 0.0042 &-0.0002 &-0.0024  \\
		    16 & $\{7,3\}$ & 0.0009 & 0.0023 &-0.0071 &-0.0026 &-0.0046  \\
		    17 & $\{7,2\}$ & 0.0001 &-0.0010 & 0.0036 &-0.0001 &-0.0007  \\
		    18 & $\{7,1\}$ & 0.0024 &-0.0007 & 0.0015 & 0.0013 &-0.0004  \\
		    19 & $\{8,4\}$ & 0      & 0      &-0.0081 & 0      &-0.0025  \\
		    20 & $\{7,0\}$ & 0.0019 &-0.0014 & 0.0012 & 0.0018 &-0.0006  \\
			\hline 
			\hline
			\hline
		\end{tabular}
\end{table}

\textcolor{black}{The spontaneous out-of-plane polarisation that we predict here for few-layer rhombohedral graphene structures is a feature induced by an asymmetrically placed twin boundary which breaks inversion and mirror symmetry in the system, and we trace it to the effect of individual terms in the full Sloczewski-Weiss-McClure model of graphite involved with such asymmetries and responsible for electron-hole symmetry breaking of the single-particle spectra. }

\textcolor{black}{The reported calculations show that the size and orientation of $P_z$ are critically affected by the intrinsic screening inside the graphitic films, leading to the opposite signs of $P_z$ in tetralayers analysed with and without screening, and much smaller values, about one order of magnitude smaller than those experimentally reported in $\mathrm{WTe_2}$ \cite{fei_ferroelectric_2018}, $\mathrm{MoS_2}$  \cite{Weston2022}or hBN \cite{Yasuda_2021}. We also note that in marginally twisted multilayers, the largest local polarisation appears to be at domain walls and the near the nodes of domain wall networks. Overall, such sensitivity to the choice of the values of input parameters in the tight-binding model of graphite, number of layers in the film and variation across the moir\'{e} pattern in marginally twisted structures may be used to experimentally refine the parametrization of Sloczewski-Weiss-McClure model.}

\begin{acknowledgement}

This work was supported by EC-FET Core 3 European Graphene Flagship Project, EC-FET Quantum Flagship Project 2D-SIPC, EPSRC grants EP/S030719/1 and EP/V007033/1, and the Lloyd Register Foundation Nanotechnology Grant.

\end{acknowledgement}

\begin{suppinfo}
In the file $``$ Supporting\_Information.pdf$"$, we describe the methods used to produce the data of the text, including:
\begin{itemize}
  \item Description of the Hamiltonian of rhombohedral graphite with a twin boundary
  \item Description of the Hamiltonian of tetralayer graphite with one interface laterally shifted by an arbitrary amount $\boldsymbol{r}_0$
  \item Analysis of the area in reciprocal space required for convergence.
  \item Description of the method used to account for electrostatic screening.
  \item Analysis of the dependence of $P_z$ and $P_z^u$ on each of the SWMcC parameters.
  \item Description of the method used to account for lattice relaxation.
\end{itemize}

\end{suppinfo}

\section*{Author contributions}
V. F. conceived the project. A. G.-R. made calculations of band structures, ferroelectric polarisation, self-consistent implemetation of screening and prepared all figures in the text. V. E. did analysis of lattice reconstruction. All the authors discussed the results and wrote the manuscript.

 \section*{Competing Interests} 
The Authors declare no competing financial or non-financial interests.

\section*{Data availability} 
The data that support the plots in the manuscript are available from the corresponding authors upon reasonable request.

\bibliography{Bibl}
\newpage

% \section{TOC Graphic}

% Ferroelectric polarisation map in marginally twisted double bilayer graphene. The energetically favourable ABAC and ABCB stacking orders, highlighted in pink and blue, become the preferential stacking order inside domains, leading to the formation of mesoscopic domain walls that intercross in regions of ABBA stacking order.

\end{document}

\section{Hamiltonian of rhombohedral graphite with a twin boundary}
To construct the Hamiltonian of rhombohedral graphite with a twin-boundary fault, we incorporate all possible couplings described in Figure 1 of the main text. In the basis of $A$ and $B$ sublattice amplitudes and from the bottommost to the topmost layers, $\{ \psi_{\beta,\boldsymbol{p}}^{A_1},\psi_{\beta,\boldsymbol{p}}^{B_1},...,\psi_{\beta,\boldsymbol{p}}^{B_{n+m+3}}\}$, this Hamiltonian takes the form
\begin{align}\label{Eq:Ham_TB}
&\mathcal{H}=
\left(
\begin{matrix}
H_\mathrm{g}^s&V&W&\cdots&0&\cdots&0&0&0\\
V^\dagger&H_\mathrm{g}^b&V&\cdots&0&\cdots&0&0&0\\
W^\dagger&V^\dagger&H^b_\mathrm{g}&\cdots&0&\cdots&0&0&0\\
\vdots&\vdots&\vdots&\ddots&\vdots&
\vdots&\vdots&\vdots&\vdots\\
0&0&0&\cdots&\mathcal{H}_{\mathrm{ABA}}&\cdots&0&0&0\\
\vdots&\vdots&\vdots&\cdots&\vdots&\ddots&\vdots&\vdots&\vdots\\
0&0&0&\cdots&0&\cdots&H^b_\mathrm{g}&V^\dagger&W^\dagger\\
0&0&0&\cdots&0&\cdots&V&H^b_\mathrm{g}&V^\dagger\\
0&0&0&\cdots&0&\cdots&W&V&H_\mathrm{g}^s
\end{matrix}
\right),\\
&H_\mathrm{g}^s=H_g+
\left(
\begin{matrix}
\Delta'&0\\
0&0
\end{matrix}
\right),
\quad
H^b_\mathrm{g}=H_g+
\Delta'\hat{1}_2
,\quad
H_g=v
\left(
\begin{matrix}
0&\pi_\xi^*\\
\pi_\xi&0
\end{matrix}
\right),\nonumber\\
&V=
\left(
\begin{matrix}
-v_4\pi_\xi&\gamma_1\\
-v_3\pi_\xi^*&-v_4\pi_\xi
\end{matrix}
\right),\quad
W=
\left(
\begin{matrix}
0&0\\
\gamma_2/2&0
\end{matrix}
\right),\quad
\pi_\xi\equiv\xi p_x+ip_y,
\nonumber\\
&\mathcal{H}_{ABA}=
\left(
\begin{matrix}
H^b_\mathrm{g}&V&\tilde{W}\\
V^{\dagger}&H^b_\mathrm{g}-\sigma_z\Delta'&V^\dagger\\
\tilde{W}^\dagger&V&H^b_{\mathrm{g}}
\end{matrix}
\right),\quad
\tilde{W}=
\left(
\begin{matrix}
\gamma_5/2&0\\
0&\gamma_2/2
\end{matrix}
\right),\nonumber
\nonumber
\end{align}
where we introduce the Sloczewski-Weiss-McClure (SWMcC) parameters\cite{McClure1957,Slonczewsi1958,McClure1960}, $v\approx1.02\cdot10^{6}$ m/s, $\gamma_1=390$ meV, $v_3\approx0.1v$, $v_4\approx0.022v$, $\Delta'=25$ meV, $\gamma_2=-17$ meV and $\gamma_5=38$ meV  \cite{yin_dimensional_2019}, and $\xi=\pm1$ is the valley index. 

\section{Hamiltonian of tetralayer graphite with a marginally twisted interface}

Local stacking characterised by an interlayer offset, $\boldsymbol{r}_0$ at the twsited interface. The following Hamiltonians generalise those derived in Ref. \cite{Garcia-Ruiz2021}. For 2AB+2BA, Hamiltonian reads
\begin{align}\label{Eq:Ham_2+2}
&\mathcal{H}_{AB+BA}=\left(
\begin{matrix}
H_\mathrm{g}^s&V&\mathcal{W}_{\mathrm{AB+1}}(\boldsymbol{r}_0)&0\\
V^\dagger&
H_\mathrm{g}^s-\sigma_z\Delta'+\mathcal{E}_b&
\mathcal{V}(\boldsymbol{r}_0)&
\mathcal{W}_{\mathrm{1+BA}}(\boldsymbol{r}_0)\\
\mathcal{W}_{\mathrm{AB+1}}^\dagger(\boldsymbol{r}_0)&
\mathcal{V}^\dagger(\boldsymbol{r}_0)&
H_\mathrm{g}^s-\sigma_z\Delta'+\mathcal{E}_t&
V^\dagger\\
0&\mathcal{W}_{\mathrm{1+BA}}^\dagger(\boldsymbol{r}_0)&
V&H_\mathrm{g}^s
\end{matrix}
\right)
\end{align}
\begin{align}
&\mathcal{V}(\boldsymbol{r}_0)=
\sum_{j=0}^2
\left\{
\left[
\frac{\gamma_1}{3}
-\frac{2v_4\hbar}{3K}
\boldsymbol{k}\cdot\boldsymbol{K}_\xi^{(j)}
\right]
\left(
\begin{matrix}
1&e^{i\xi\frac{2\pi}{3}j}\\
e^{-i\xi\frac{2\pi}{3}j}&1
\end{matrix}
\right)\right.\nonumber\\
&+\left.
\frac{\xi2(v_3-v_4)\hbar}{3K}
\left[
\boldsymbol{k}\times\boldsymbol{K}_\xi^{(j)}
\right]_z
\left(
\begin{matrix}
0&ie^{i\xi\frac{2\pi}{3}j}\\
-ie^{-i\xi\frac{2\pi}{3}j}&0
\end{matrix}
\right)
\right\}e^{-i\boldsymbol{K}_\xi^{(j)}\cdot
\boldsymbol{r}_0},\nonumber\\
&\mathcal{W}_{\mathrm{AB+1}}
(\boldsymbol{r}_0)=
\frac{1}{6}
\sum_{j=0}^2
\left(
\begin{matrix}
\gamma_5e^{-i\xi\frac{2\pi}{3}j}&
\gamma_5\\
\gamma_2e^{+i\xi\frac{2\pi}{3}j}&
\gamma_2e^{-i\xi\frac{2\pi}{3}j}
\end{matrix}
\right)
e^{-i\boldsymbol{K}_\xi^{(j)} \cdot \boldsymbol{r}_0},\nonumber\\
&\mathcal{W}_{\mathrm{1+BA}}
(\boldsymbol{r}_0)=
\frac{1}{6}
\sum_{j=0}^2
\left(
\begin{matrix}
\gamma_5e^{+i\xi\frac{2\pi}{3}j}&
\gamma_2e^{-i\xi\frac{2\pi}{3}j}\\
\gamma_5&
\gamma_2e^{+i\xi\frac{2\pi}{3}j}
\end{matrix}
\right)
e^{-i\boldsymbol{K}_\xi^{(j)} \cdot \boldsymbol{r}_0},\nonumber\\
%%%%%%%%%%%%%
%%%%%%%%%%%%%
&\mathcal{E}_{b/t}=
\frac{2\Delta'}{3}
\hat{1}_2+
\frac{\Delta'}{9}
\sum_{\boldsymbol{G}}
e^{i\boldsymbol{G}\cdot\boldsymbol{r}_0}
\left(
\begin{matrix}
1+e^{\pm i\boldsymbol{G}\cdot\boldsymbol{\tau}_B}&0\\
0&1+e^{\mp i\boldsymbol{G}\cdot\boldsymbol{\tau}_B}
\end{matrix}
\right),\nonumber
\end{align}
where we introduce K-valleys momenta, $\boldsymbol{K}_\xi^{(j)}=\xi K[\cos(j2\pi/3),-\sin(j2\pi/3)]$, with $K=4\pi/3a$ ($a\approx2.46$ {\AA} is the lattice constant of graphene). We note that for $\boldsymbol{r}_0=(0,a/\sqrt{3})$ and $\boldsymbol{r}_0=(0,-a/\sqrt{3})$, Eq. \eqref{Eq:Ham_2+2} transforms to the Hamiltonian of ABAC and ABCB, respectively. Likewise, the Hamiltonian of monolayer-twisted-trilayer structures is given by
\begin{subequations}\label{Eq:Ham_3+1}
\begin{align}
&\mathcal{H}_{\mathrm{ABA+1}}=\left(
\begin{matrix}
H_\mathrm{g}^s&V&\tilde{W}&0\\
V^\dagger&
H^b_\mathrm{g}-\sigma_z\Delta'&
V^\dagger&
\mathcal{W}_{\mathrm{BA+1}}(\boldsymbol{r}_0)\\
\tilde{W}^\dagger&V&
H_\mathrm{g}^s+\mathcal{E}_b&\mathcal{V}(\boldsymbol{r}_0)
\\
0&\mathcal{W}_{\mathrm{BA+1}}^{\dagger}&
\mathcal{V}^\dagger(\boldsymbol{r}_0)&H_\mathrm{g}+\mathcal{E}_t
\end{matrix}
\right),\\
&\mathcal{H}_{\mathrm{ABC+1}}=
\left(
\begin{matrix}
H_\mathrm{g}^s&V&W&0\\
V^\dagger&
H^b_\mathrm{g}&
V&
\mathcal{W}_{\mathrm{AB+1}}
(\boldsymbol{r}_0)\\
{W}^\dagger&V&
H_\mathrm{g}^s-\sigma_z\Delta'+\mathcal{E}_b&
\mathcal{V}(\boldsymbol{r}_0)
\\
0&\mathcal{W}_{\mathrm{AB+1}}^{\dagger}(\boldsymbol{r}_0)&
\mathcal{V}^\dagger(\boldsymbol{r}_0)&H_\mathrm{g}+\mathcal{E}_t
\end{matrix}
\right),\\
&\mathcal{W}_{\mathrm{BA+1}}
(\boldsymbol{r}_0)=
\frac{1}{6}
\sum_{j=0}^2
\left(
\begin{matrix}
\gamma_2e^{+i\xi\frac{2\pi}{3}j}&
\gamma_2e^{-i\xi\frac{2\pi}{3}j}\\
\gamma_5&
\gamma_5e^{+i\xi\frac{2\pi}{3}j}
\end{matrix}
\right)
e^{-i\boldsymbol{K}_\xi^{(j)} \cdot \boldsymbol{r}_0}.\nonumber
\end{align}
\end{subequations}

\section{Convergence analysis}
\begin{figure}
\begin{center}
\includegraphics[width=1\columnwidth]{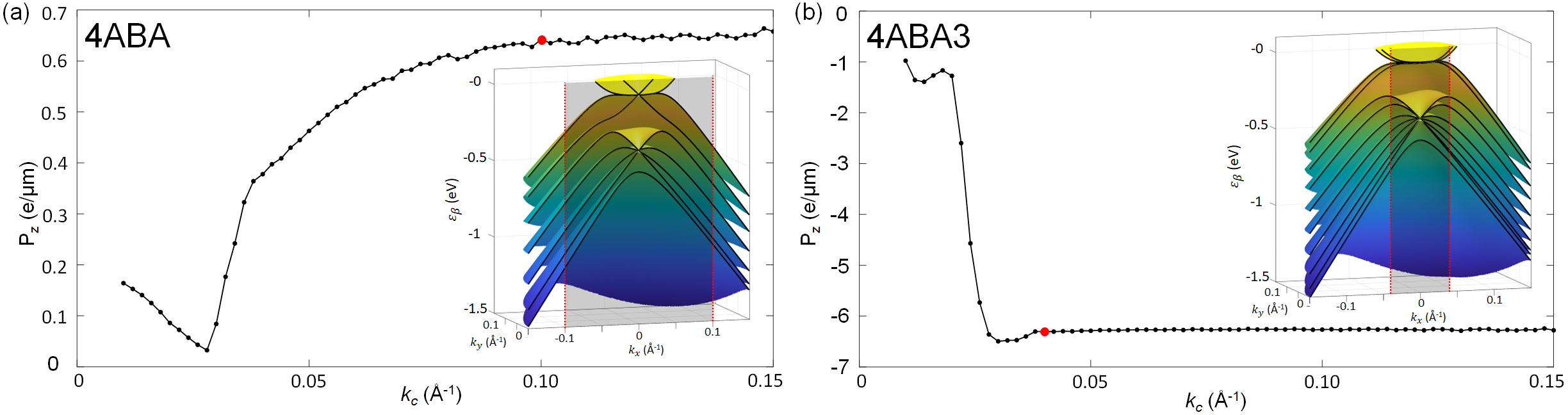}
\caption{ Value for the ferroelectric polarisation, $P_z$, as a function of the radius of the circle centred at $\boldsymbol{K}_{\pm}$ in reciprocal space, $k_c$ taken to compute the electron density in each layer, for 4ABA (a) and 4ABA3 (b) tetralayer graphene. For the latter, it suffices to sample the area on which flat bands extend to obtained a converged value.
\label{Fig_convergence}}
\end{center}
\end{figure}

In our analysis, we compute the electron density in each layer by brute-force diagonalisation of the Hamiltonian \eqref{Eq:Ham_TB} for a sufficiently dense grid of points in reciprocal space within a circular area with cut-off wavevector, $k_c$. For each nABAm film, as $k_c$ increases, the ferroelectric polarisation converges to a $P_z$-value, displayed below in the left panel of Fig. \ref{Fig:self-consistency} (c). To illustrate that this value of the cut-off is large enough, we consider behaviour of $P_z $ versus $k_c$ for two characteristic structures: 4ABA and 4ABA3, shown in Fig. \ref{Fig_convergence}. 

The low-energy dispersion of 4ABA graphite film features a couple of flat bands and a pair of Dirac dispersive bands [see inset in Fig. \ref{Fig_convergence} (a)]. The former, strongly localised at the bottom and twin boundary layers, induces a polarisation in the positive $z$-direction, while the latter, hybridising with deeper bands, contributed to $P_z$ in negative direction of $z$-axis. As shown Fig. \ref{Fig_convergence} (a), it is necessary to sample an area of $5\cdot10^{-2}$ {\AA}$^{-2}$ (or a value for the cut-off energy of $\sim-0.13$ eV) to reach convergence. 

In contrast, the low-energy dispersion of 4ABA3 exhibits two pairs of flat bands [see inset in Fig. \ref{Fig_convergence} (b)], with a much stronger localisation at the twin boundary and non-dimer site of surface layer, that are well-separated from the rest continuum of the deep-energy states. In this case $P_z$ converges at smaller area of reciprocal space, about $5\cdot10^{-3}$ {\AA}$^{-2}$ (or a value for the cut-off energy of $\sim-0.07$ eV), which corresponds to the region of high density of states of the flat bands. 

\section{Self-consistent implementation of screening}
Here, we extend the above-described SWMcC model to incorporate the effects of internal electric fields produced by the charge redistribution. These fields produce an energy offset in each layer, which can be captured in our model by adding a term $\epsilon_i \hat{1}_2$ to the i-th diagonal block of the Hamiltonians in Eqs. (\ref{Eq:Ham_TB}), (\ref{Eq:Ham_2+2}) and (\ref{Eq:Ham_3+1}) using the formula
\cite{Slizovskiy_dielectric_2021}
\begin{align}\label{Eq:Ediffs}
\epsilon_i-\epsilon_{i-1}=
    \frac{e^2d}{2\varepsilon_0}
    \left[
    (n_i-n_{i-1})
    \frac{1+\varepsilon_z^{-1}}{2}+
    \sum_{j>i}
    {n_j}{\varepsilon_z^{-1}}-
    \sum_{j'<i-1}
    {n_j'}{\varepsilon_z^{-1}}
    \right],
\end{align}
with $d\approx3.35$ \AA~ being the interlayer distance in graphene stacks, $\varepsilon_z\approx2.6$\cite{Slizovskiy_dielectric_2021} the relative permittivity of graphene. We determine the Fermi level for neutral structures, and calculate the charge redistribution across the layers, taking into account only the eigenvectors associated to states below the Fermi level. Because the charge densities $n_i$ also depend implicitly on the on-layer potentials, the expression in Eq. (\ref{Eq:Ediffs}) needs to be solved self-consistently. In the first iteration, we set all on-layer energies to zero, $\epsilon_i^{(0)}=0$, which gives the value of polarisation for the unscreened system, $P_z^u$. In each iteration, we use Eq. (\ref{Eq:Ediffs}) to compute a new set of on-layer energies, $\tilde{\epsilon}_i^{(n)}$, which are used to update the value employed in the following iteration using
\begin{align*}
    \epsilon_i^n=\eta \tilde{\epsilon}_i^{(n)}+(1-\eta)\epsilon_i^{(n-1)},
\end{align*}
where $\eta$ determines the fraction of the value for the on-layer energies that is updated in each iteration, and was taken to be $4\cdot10^{-4}$ in our work. This linear-mixing algorithm is necessary in a broad range of self-consistent methods to avoid overshooting \cite{Woods_Computing_2019}.

\begin{figure}
\begin{center}
\includegraphics[width=0.75\columnwidth]{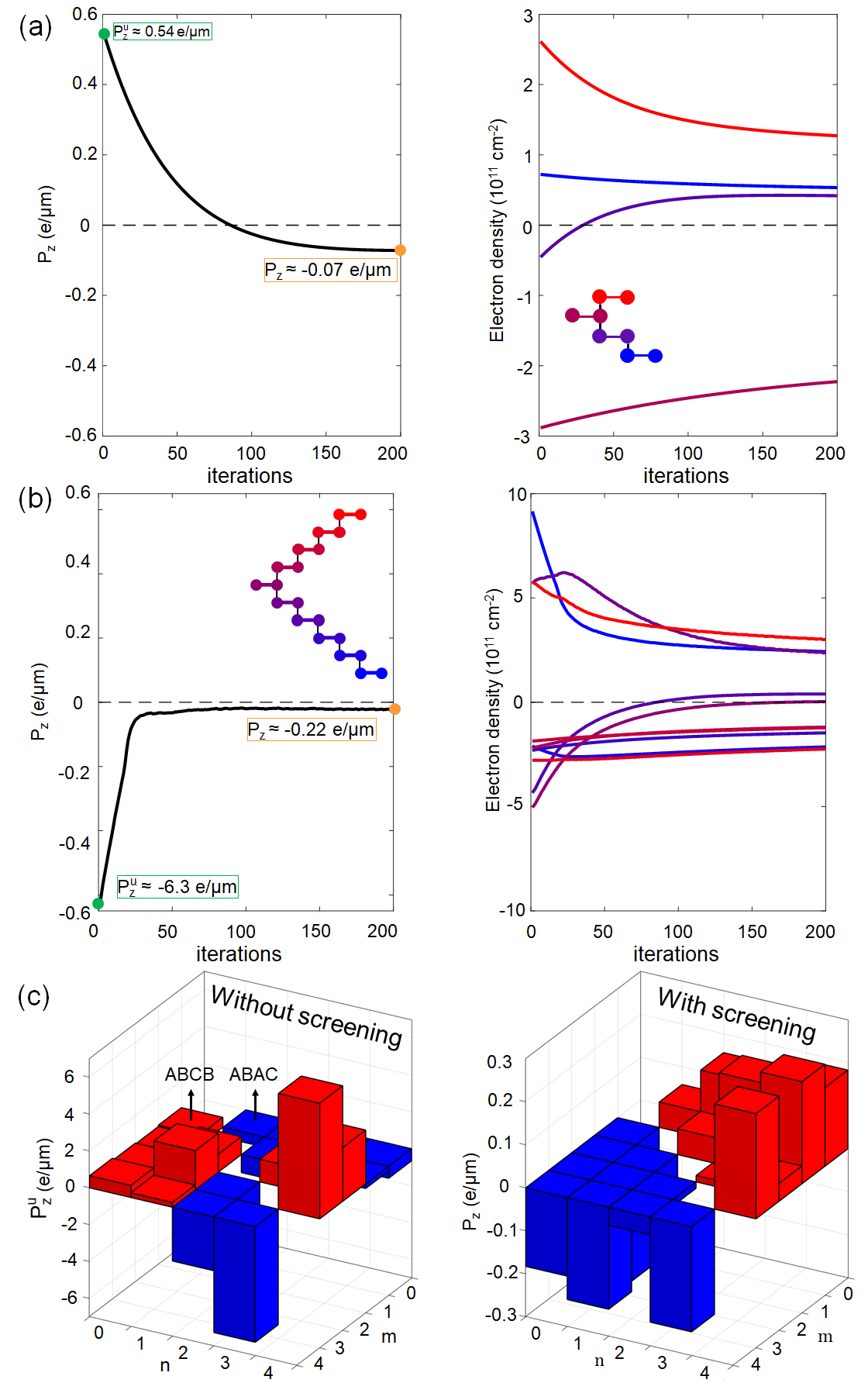}
\caption{ Convergence analysis of $P_z$ and $n_i$ for (a) 1ABA and (b) 4ABA3 films. The latter evidences that, for larger films, electron charge primarily redistributes to the surface and twin boundary layers. (c) Electric polarisation of mABAn, $P_z^u$ ($P_z$), displayed in the form of a histogram, using a model that neglects (accounts for) screening effects of electron-density redistribution. 
\label{Fig:self-consistency}}
\end{center}
\end{figure}

In Figs. \ref{Fig:self-consistency}(a) and \ref{Fig:self-consistency}(b), we show the value of $P_z$ and electron densities in each layer as a function of the iteration for 1ABA and 4ABA3 films, highlighting importance of self-consistent calculations. Histograms in Fig. \ref{Fig:self-consistency}(c) show comparison of $P_z$ values for mABAn films with less than 8 layers computed with (right) and without (left) account of self-consistent screening. We also note that our results strongly depend on the input parameters, and suggest that measuring $P_z$ in these systems could be an experimental route to narrow down the possible values of the SWMcC parameters.

\section{Analysis of parametric dependence}

\begin{figure}
\begin{center}
\includegraphics[width=1\columnwidth]{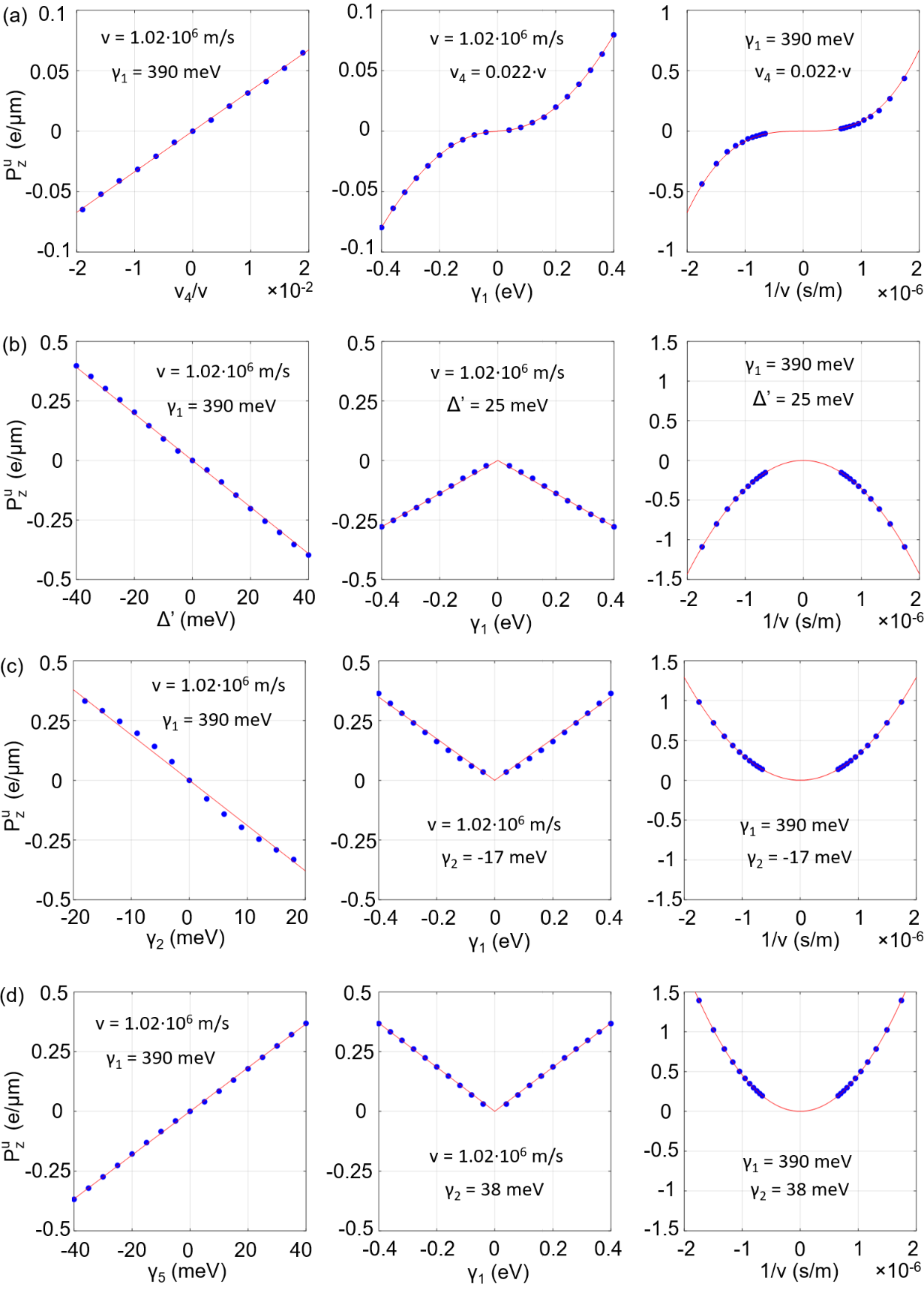}
\caption{ Numerical values (blue points) and fitting curves (red) computed to support the parametric dependence of the first, second, third and fourth terms in Eq. (2) in the main text are analysed in the triad of panels (a), (b), (c) and (d), respectively, for ABCB tetralayer graphene. In the leftmost panels, we demonstrate the linear dependence of each SWMcC parameter that induces ferroelectric polarisation. In the middle (rightmost) panel, we analyse the $\gamma_1|\gamma_1|$ and $|\gamma_1|$ ($v^{-3}$ and $v^{-2}$) dependence of the first and the remaining three terms in Eq. (2) of the main text, respectively. Parameters not shown in each graph were taken to be zero.
\label{Fig:parametric_analysis}}
\end{center}
\end{figure}

\begin{figure}
\begin{center}
\includegraphics[width=1\columnwidth]{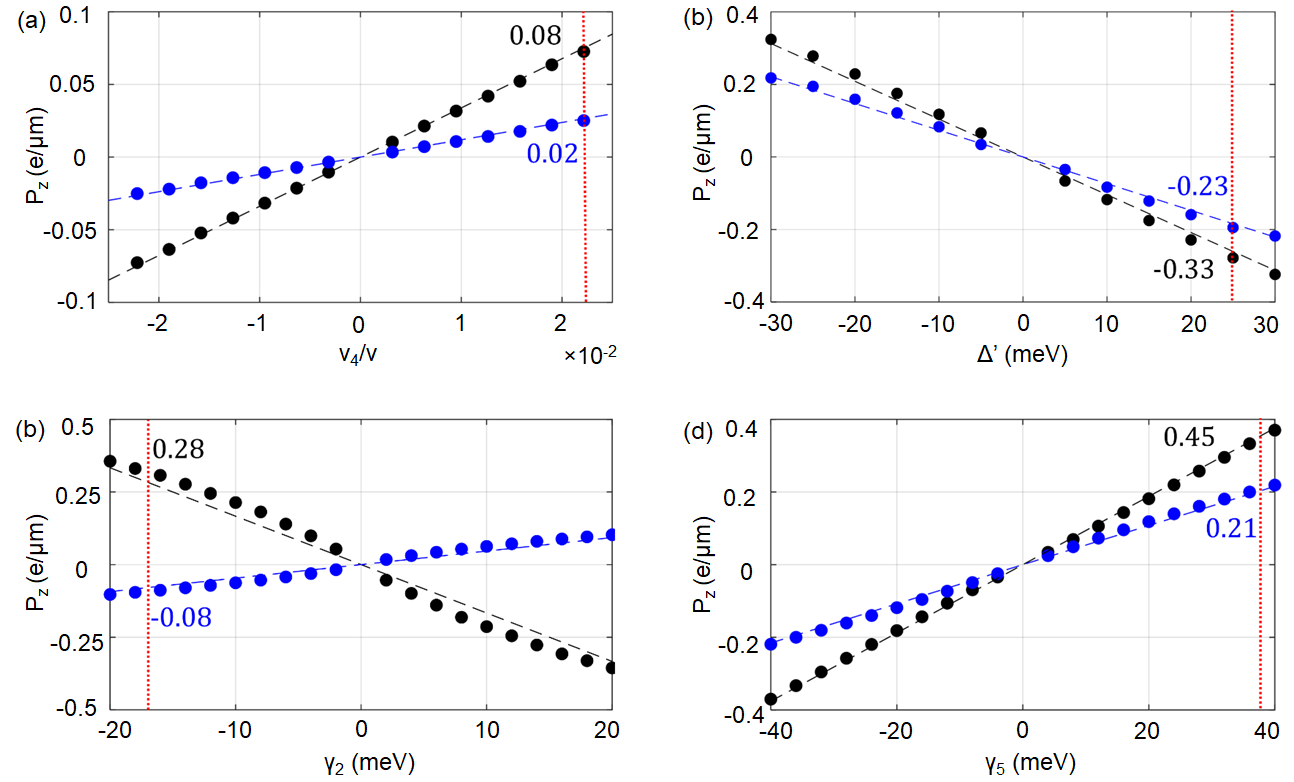}
\caption{ Dependence of $P_z^u$ (black) and $P_z$ (blue) as a function of each individual SWMcC parameter responsible for electron-hole asymmetry. Vertical red dotted lines lie on the values given in \cite{yin_dimensional_2019}, which we use in our work. Adding all the values where they intersect the black and blue interpolating lines amounts approximately the value por $P_z^u$ and $P_z$ obtained using the full SWMcC model, and shown as the first and last values in Fig. \ref{Fig:self-consistency}(a).
\label{Fig:parametric_analysis_comparison}}
\end{center}
\end{figure}
Equation (3) of the main text is based on the analysis shown in Fig. \ref{Fig:parametric_analysis}, where each panel represents dependence of $P_z^u$ of ABCB tetralayer graphene on one of the SMWcC parameters. The blue dots are numerically computed values of $P_z^u$ and the red lines are the linear interpolations, which we use to obtain the coefficients $\mathcal{X}_{4,D,2,5}$. We observe the linear dependence of $P_z$ on the symmetry-breaking parameters $v_4$, $\Delta'$, $\gamma_2$ and $\gamma_5$. The middle and rightmost panels of Fig. \ref{Fig:parametric_analysis} (a), shows the dependence on $\gamma_1$ and $v$, when $v_4$ is kept as the only non-zero electron-hole symmetry-breaking parameter, which is proportional to $\gamma_1|\gamma_1| \times v^{-3}$, respectively. Likewise, in the same panels of Figs. \ref{Fig:parametric_analysis} (b), (c) and (d), we demonstrate that $P_z^u$ as the power law $|\gamma_1|\times v^{-2}$ when $\Delta'$, $\gamma_2$ and $\gamma_5$ are kept as the only non-zero parameters, respectively. 

In Fig. \ref{Fig:parametric_analysis_comparison}, we extend  analysis of $P_z$ when effect of self-consistent screening is taken into account. In particular, we compare the leftmost panels of Fig. \ref{Fig:parametric_analysis} using a model with (blue dots) and without (black dots) the effects of screening, and demonstrate that, while $\mathcal{X}_4$, $\mathcal{X}_D$ and $\mathcal{X}_5$ reduce their values by a factor of $\sim$4, 1.4 and 2.2, respectively, the coefficient $\mathcal{X}_2$ changes sign. As stated in the main text, this peculiar behaviour plays a prominent role in the change of sign of polarisation, being $P_z^u\approx0.54~\mathrm{e/\mu m}$ in the unscreened model and $P_z\approx-0.07 ~\mathrm{e/\mu m}$ in the screened model [see Fig. \ref{Fig:self-consistency}(a)]. It is also worth mentioning that adding the individual contributions to $P_z$ (or $P_z^u$) from the four symmetry-breaking terms amounts approximately to the to computed using the full SWMcC model with (without) the account for screening effects.

\section{Lattice relaxation in twisted tetralayer structures}

To describe lattice reconstruction of moir\'e superlattice formed at the twisted interface we introduce in-plane displacement fields $\bm{u}_{t/b}$ acting in top/bottom stacks. For 3+1 tetralayers top (bottom) stack consists of 1\,(3) graphene layers, while for 2+2 structures the top/bottom stacks are graphene bilayers. To find $\bm{u}_{t/b}$ we minimize sum of elastic, $U=\sum_{l={t,b}}\left[(\lambda_l/2)\left(u_{ii}^{(l)}\right)^{2} + \mu_l u_{ij}^{(l)}u_{ji}^{(l)}\right]$, and local adhesion, $W\left(\bm{r}_0=\theta \hat{z}\times\bm{r}+\bm{u}^{(t)}-\bm{u}^{(b)}\right)$, energies over supercell with periodicity of moir\'e superlattice given by twist angle $\theta$ expanding the displacement fields into Fourier series \cite{EnaldievPRL2020}: $\bm{u}_{t/b}=\sum_{\{\lambda\},j=0,1,2}\left[\hat{R}_{2\pi j/3}\bm{u}^{(t/b)}_{\{\lambda\}}\right]\sin\left(\bm{g}_{\{\lambda\}}^{(j,+)}\bm{r}\right)$ with $\bm{g}_{\{\lambda\}}^{(j,+)}=\theta\bm{G}_{\{\lambda\}}^{(j,+)}\times\hat{z}$ and $\hat{R}_{\phi}$ is counter clock-wise rotation on angle $\phi$ around $\hat{z}$. Obtained values of non-negligible Fourier amplitutes are listed in Table \ref{Tab:utb}. In calculations, elastic moduli of top/bottom stacks were determined by those of graphene $\lambda_{G}=c_0Y\nu/(1+\nu)(1-2\nu)$, $\mu_{G}=c_0Y/2(1+\nu)$ (with Young module, $Y=1$\,TPa, and Poisson ratio, $\nu=0.19$ \cite{androulidakis2018tailoring}), multiplied by the number of graphene layers in each stack. For the adhesion energy at twisted interface we used the following expression \cite{SrolovitzPRB2015,carr2018}:   $W(\bm{r}_0)=\sum_{\Lambda=1,2,3}w_\Lambda\sum_{j=0,1,2}\cos\left(\bm{G}^{(j,+)}_{\Lambda},\bm{r}_0\right)$ with $w_1=0.775{\rm\,meV/A^2}$, $w_2=-0.071{\rm\,meV/A^2}$, $w_3=-0.018{\rm\,meV/A^2}$. 

\begin{table}
		\caption{Main non-zero Fourier series coefficients (in $\rm{\AA}$) of displacement fields describing reconstruction in twistronic tetralayers 3ABA+1, 3ABC+1 and 2AB+2BA structures with $\theta=0.05^{\circ}$, which were used in Fig. 2 of the main manuscript.}
		{%\small
			\begin{tabular}{c|c|c||c}%c|c|c|c|c|c|c|}
				%\hline
				\hline
				\hline
					%${\lambda_1,\lambda_2}$ & \multicolumn{3}{c|}{P} & \multicolumn{2}{c}{AP} \\
                    & \multicolumn{2}{c||}{3AB(A/C)+1} & 2AB+2BA \\
                    \hline
                 	$\{\lambda_1,\lambda_2\}$ & (1L) $u_{\lambda_1,\lambda_2}^{(t)}$ & (3L) $u_{\lambda_1,\lambda_2}^{(b)}$ & $u^{t}_{\lambda_1,\lambda_2}(=-u^{b}_{\lambda_1,\lambda_2})$ \\
                  \hline
                  1, 0 & (0.2898, 0.1687)  & $-$(0.0979, 0.0565)  & (0.1938, 0.1119) \\
                  2, 0 & (0.1394, 0.0797)  & $-$(0.0463, 0.0265)  & (0.0910, 0.0526) \\
                  3, 0 & (0.0875, 0.5089)  & $-$(0.0292, 0.0169)  & (0.0573, 0.0330) \\
                  4, 0 & (0.0606, 0.0349)  & $-$(0.0202, 0.0116)  & (0.0389, 0.0225) \\
                  5, 0 & (0.0428, 0.0247)  & $-$(0.0143, 0.0082)  & (0.0264, 0.0153) \\
                  6, 0 &  (0.0322, 0.0185) & $-$(0.0107, 0.0062)  & (0.0201, 0.0116) \\
                  7, 0 &  (0.0227, 0.0132) & $-$(0.0076, 0.0044)  & (0.0130, 0.0075) \\
                  8, 0 &  (0.0191, 0.0109) & $-$(0.0064, 0.0037)  & (0.0121, 0.0070) \\
				\hline
				\hline
			\end{tabular}
			\label{Tab:utb}}
	\end{table}

\newpage
\bibliography{Bibl}